\documentclass[aps,pra,twocolumn,groupedaddress,showpacs,superscriptaddress,scrartcl,longbibliography]{revtex4-2}
\usepackage{bbold,array,xcolor,enumitem,tabularx,multirow,hyperref,comment,amssymb,amsmath,graphicx,epstopdf,amsthm,dsfont,enumerate,bm,color}
\newtheorem{theorem}{Theorem}

\newtheorem{criterion}{Criterion}

\newtheorem{assumption}{Assumption}

\newtheorem{definition}{Definition}

\newtheorem*{lemma1*}{Lemma 1}
\newtheorem*{lemma2*}{Lemma 2}
\newtheorem*{lemma3*}{Lemma 3}
\newtheorem*{lemma4*}{Lemma 4}
\newtheorem*{lemma5*}{Lemma 5}
\newtheorem*{lemma8*}{Lemma 8}
\newtheorem*{lemma9*}{Lemma 9}
\newtheorem*{lemma10*}{Lemma 10}
\newtheorem*{lemma11*}{Lemma 11}
\newtheorem*{theorem1*}{Theorem 1}

\newtheorem*{theorem2*}{Theorem 2}
\newtheorem*{theorem3*}{Theorem 3}

\newcommand{\edit}[1]{{\color{black}{#1}}}

\begin{document}

\title{Is spacetime quantum?}

\author{Dami\'an Pital\'ua-Garc\'ia}
\email{D.Pitalua-Garcia@damtp.cam.ac.uk}
\affiliation{Centre for Quantum Information and Foundations, DAMTP, Centre for Mathematical Sciences, University of Cambridge, Wilberforce Road, Cambridge, CB3 0WA, U.K.}

\date{\today}

\begin{abstract}

Although the standard viewpoint in theoretical physics is that the unification of quantum theory and general relativity requires the quantization of gravity and spacetime, there is not consensus about whether spacetime must fundamentally have any quantum features. 
Here we show a theorem stating that spacetime degrees of freedom \edit{and a quantum system violate a Bell inequality} in a background Minkowski spacetime if a few properties of general relativity and quantum theory have a broad range of validity, and if the quantum state reduction upon measurement is a real physical process that is completed superluminally when acting on distant quantum particles in a quantum entangled state. \edit{We argue that this implies that spacetime cannot be sensibly called classical if the assumptions in our theorem hold.} In contrast to the Eppley-Hannah argument for the necessity of quantizing the gravitational field, we discuss the validity of our assumptions, our thought experiment does not require to manipulate or detect gravitational waves, and our theorem does not rely on the conservation of momentum or on the uncertainty principle.

\end{abstract}

\maketitle

\section{Introduction}



Quantum theory (QT) and general relativity (GR) are our most fundamental theories of physics. They are both supported by enoumous experimental confirmation for over a century. 
However, QT and GR cannot provide a
complete description of physics. On the one hand, QT predicts that the configuration of matter and energy can be in \emph{quantum superpositions}. On the other hand, Einstein's equations of GR treat matter, energy and spacetime as purely \emph{classical}, which cannot be in quantum superpositions. 
The prediction of QT that massive systems can be in quantum superpositions has been experimentally confirmed in microscopic scales (e.g., \cite{ANVKVGZ99,CSP09,HGHNA12,EGAMT13,AH14,FGZKPMGA19}). Thus, it follows that GR needs to be modified at least in these scales. Moreover, it is conceivable that QT requires modification in sufficiently macroscopic scales too. Importantly, it has been suggested that the predictions of QT could reach a limit of validity at sufficiently macroscopic scales due to effects of gravity (e.g., \cite{K66,D87,D89,P96,P98,BLSSU13,BGU17}).

This motivates the question, \emph{is spacetime fundamentally classical, or does it have some quantum properties?} 
It could be argued that because QT has been experimentally confirmed in microscopic scales, spacetime must have some quantum properties in these scales too. In fact, several theories of quantum gravity have been investigated (e.g., string theory \cite{GreenSchwarzWittenbook}, loop quantum gravity \cite{Rovellibook}, and others \cite{H05,K06,Oritibook,H18}). Furthermore, there are arguments claiming that 
the gravitational field must be quantized (e.g., \cite{DR11,EH77,T06,MV17.2}).
However, these arguments have been refuted (e.g., \cite{M06,AKR08,K18,O18}).
Moreover, quantizing the gravitational field
has several important problems \cite{C01,Keifer13}. Thus, it has been proposed that 
the gravitational field could be fundamentally classical (e.g., \cite{M62,R63,HC01,M05,W05,C08,K18,O18}).  

\edit{In particular, semi-classical gravity models \cite{M62,R63} assume that the spacetime metric and the gravitational field are classical and sourced by the expectation value of the stress-energy tensor of quantum matter propagating in spacetime according to Einstein's equations of GR. Crucially, semi-classical gravity precicts that matter can be in quantum superpositions, while the generated gravitational field and the spacetime geometry are not. For example, the location of a macroscopic massive object can be in a quantum superposition, while the generated gravitational field is in a classical state given by the expectation value of the quantum state. More broadly, it is in principle possible that gravity is described by a non-quantum theory -- not necessarily semi-classical gravity -- predicting that matter can be in quantum superpositions while the gravitational field remains classical (e.g., \cite{M62,R63,HC01,M05,W05,C08,K18,O18}).

It is crucial to have the previous point clear to follow our arguments in this paper. We present below a thought experiment in which a macroscopic massive system can be set in a quantum superposition. It follows from the previous discussion that this does not imply that the gravitational field evolves into a quantum superposition too. Otherwise, the quantum nature of gravity would follow trivially.

As discussed above,} there is not consensus at present of whether the gravitational field, and hence spacetime, must have any quantum properties.
Ultimately, this question must be answered experimentally.
But experimental evidence for quantum features of gravity remains missing.


Different criteria could be considered to evaluate whether spacetime is fundamentally classical.
For example, natural conditions to impose on spacetime degrees of freedom to consider spacetime as classical could be: that they cannot be in quantum superpositions, or that they cannot be entangled with other systems, or that they cannot violate any Bell inequalities.

Bose \emph{et al}. \cite{BMMUTPGBKM17}, and Marletto and Vedral \cite{MV17} have recently proposed a beautiful experiment that they claim could be implemented in the near future and that could demonstrate that the gravitational field can mediate entanglement
, implying that the gravitational field is non-classical \edit{under some assumptions} \cite{BMMUTPGBKM17,MV17}. 
Assuming that the gravitational field is a property of spacetime, as GR has taught us, this would imply that spacetime is non-classical too.



Quantum entanglement is a paramount phenomenon that does not arise in classical physics. 
It has various fundamental applications in quantum information science, for example, in quantum teleportation \cite{teleportation}, superdense coding \cite{sdc}, quantum cryptography \cite{E91} and quantum computation \cite{J97}. However, the entangled correlations between a pair of quantum systems arising in timelike separated experiments can in principle be modelled with purely classical intuitions via local hidden variable theories (LHVTs) \cite{Bell}. For example, two parties communicating subluminally can straightforwardly simulate
any quantum entangled correlations.




However, as follows from Bell's theorem \cite{Bell}, the violation of Bell inequalities in experiments performed on a pair of physical systems at spacelike separation cannot be described by LHVTs, defying all classical intuitions. \edit{Because only entangled states can violate Bell inequalities \cite{BCPSW14}, the violation of a Bell inequality in a Bell experiment provides a way to witness entanglement}. QT predicts the violation of Bell inequalities by some entangled states: every pure entangled quantum state violates a Bell inequality \cite{G91,YCZLO12}, but not every mixed entangled quantum state does \cite{W89}.

\edit{From the previous points, since the Bose \emph{et al}.-Marletto-Vedral (BMV) experiment does not impose any spacetime constraints, any entangled correlations observed in the experiment can be described by LHVTs \cite{KPG21}. However, the entanglement generated by gravity (if it does) in the BMV experiment can be verified by the violation of a Bell inequality in a Bell experiment applied to sufficiently distant quantum particles to which the gravity-induced entanglement is transferred, as proposed in the Bell-Bose \emph{et al}.-Marletto-Vedral (BBMV) experiment \cite{KPG21}.}

The Clauser-Horne-Shimony-Holt (CHSH) inequality \cite{CHSH69} is one of the simplest and most popular Bell inequalities. It involves Alice randomly choosing and applying one of two measurements labelled by $\alpha\in\{0,1\}$ on a system $A$ at spacelike separation from Bob randomly choosing and applying one of two measurements labelled by $\beta\in\{0,1\}$. Alice's and Bob's respective outcomes $a$ and $b$ have two posible values $a,b\in\{0,1\}$. The CHSH inequality is given by
\begin{equation}
\label{chsh}
\lvert I_{\text{CHSH}} \rvert \leq 2,
\end{equation}
where $I_{\text{CHSH}}\equiv E(0,0)+E(0,1)-E(1,0)+E(1,1)$,
and where $E(\alpha,\beta)=P(a=b\vert\alpha,\beta)-P(a\neq b\vert\alpha,\beta)$
denotes the expectation value when Alice and Bob choose the measurements labelled by $\alpha$ and $\beta$, for all $a,b,\alpha,\beta\in\{0,1\}$. The CHSH inequality is maximally violated by a pair of qubits in a maximally entangled state, for example in the singlet state
\begin{equation}
\label{e0}
\lvert\Psi^-\rangle=\frac{1}{\sqrt{2}}\bigl(\lvert 0\rangle\lvert1\rangle-\lvert 1\rangle\lvert0\rangle\bigr),
\end{equation}
achieving the quantum Tsirelson bound \cite{T87}:
\begin{equation}
\label{cirel}
\lvert I_{\text{CHSH}} \rvert \leq 2\sqrt{2}.
\end{equation}

Before 2015, the violation of Bell inequalities had been confirmed in various experiments (e.g. \cite{FC72,AGR81,AGR82,ADR82,WJSWZ98,RKMSIMW01,MMMOM08,AWBHLNOSWWCM09,SUKRMHRFLJZ10,GMRWKBLCGNUZ13}), but subject to at least one of three important loopholes: the locality loophole \cite{Bell}, the freedom-of-choice loophole \cite{Bell} and the detection loophole \cite{Pearle70}. Three outstanding experiments \cite{HBDRKBRVSAAPMMTEWTH15,GVMWHHPSKLAAPMBGLSNSUWZ15,SMCBWSGGHACDHLVLTMZSAAPJMKBMKN15} published in 2015 demonstrated the violation of Bell inequalities closing these loopholes simultaneously. However, despite being addressed in an experiment \cite{SBHGZ08}, the collapse-locality loophole \cite{K05} remains open \cite{K20}. Thus, the results in previous Bell experiments could in principle be described by LHVTs. 

In this paper we \edit{argue} that the \emph{loophole-free} satisfaction of Bell inequalities by spacetime degrees of freedom in \edit{Bell experiments with the required spacelike separations} is a necessary condition for spacetime to be sensibly called ``classical". As we will show, under some assumptions, \edit{a bipartite system comprising a quantum system and} spacetime degrees of freedom can violate the CHSH inequality, implying that spacetime has quantum features.



Another fundamental problem of QT is the quantum measurement problem. 
As presently understood, there are two general types of quantum evolution: the continuous unitary evolution (U) of a closed system described by the Schr\"{o}dinger equation, and the apparently discontinuous quantum state reduction (QSR) upon a quantum measurement. In a way, the quantum measurement problem is to understand the interrelation between QSR and U. Since a quantum measurement comprises an interaction between the measurement apparatus and the measured system, we would expect that QSR and U should have the same fundamental description.

Different approaches towards the quantum measurement problem have been proposed. An important viewpoint is that QSR is merely a Bayesian updating of a probability distribution on acquiring new information (e.g., \cite{B70,CFS02,S07,HS10,FS14}). 

\edit{The} ``many worlds'' interpretations of QT \cite{E57,DeWittbook} claim that only U takes place fundamentally, and that the different possible outcomes in a quantum measurement coexist in reality. However, these interpretations cannot effectively explain the observed probabilities in agreement with the Born rule in quantum measurements, and they cannot be verified or refuted in experiment either \cite{K10}.

In our view, promising approaches to solve the quantum measurement problem are collapse models, proposing that QSR is a real physical process whose explanation requires to extend the Schr\"{o}dinger equation (e.g., \cite{P76,G84,GRW85,GRW86,D88,G89,GPR90,P99,P05,Tumulka06,B11,BLSSU13}). Crucially, we note that there are collapse models suggesting that gravity mediates QSR (e.g., \cite{K66,D87,D89,P96,P98,BLSSU13}).

Assuming that QSR is a real physical process, Eppley and Hannah claimed in an influential paper \cite{EH77} that the gravitational field must be quantized. Their argument is that 1) if a gravitational wave of arbitrarily small momentum can be used to reduce the quantum state of a quantum particle to measure its position then either momentum is not conserved or the uncertainty principle is violated; and 2) if the gravitational wave does not reduce the quantum state of the particle then a superluminal signal can be transmitted by having the wave interact with a particle that is entangled with another distant particle. This argument has been refuted on different grounds (e.g., \cite{HC01,M06,AKR08,K18,MOS20}). Ref. \cite{M06} shows that the device proposed by Eppley and Hannah to measure the position of a particle with a gravitational wave cannot be built even in principle. Refs. \cite{AKR08,K18} assert that even if QSR is a real physical process, it does not need to allow superluminal signalling. Refs. \cite{HC01,MOS20} argue that strict conservation of momentum does not need to hold fundamentally.

Broadly speaking, this paper presents a theorem stating that if the predictions of GR are valid for mass distributions in classical states at sufficiently small scales, the predictions of QT are universally valid, and QSR 
is a real physical process, then there \edit{is a bipartite system comprising a quantum system and} spacetime degrees of freedom violating a Bell inequality.
Thus, \edit{we argue that}, given our assumptions, spacetime must have some quantum features.


Assuming that QT has universal validity is very strong, as it has only been experimentally tested in microscopic scales, and it is conceivable that its predictions will fail in experiments in sufficiently macroscopic scales. 
Similarly, it could be the case that GR is not valid in all situations, even for mass distributions in classical states. However, our theorem only requires to assume a few properties about QSR, GR and QT, specific to a thought experiment on which our argument is based. In particular, we only need to assume a couple of rather weak properties about GR. Moreover, we do not need to assume that QT is valid in all scales. 

In contrast to the Eppley-Hannah argument \cite{EH77}, we give our assumptions explicitly and we discuss their validity. 
Crucially, our thought experiment does not require to manipulate or detect gravitational waves. Furthermore, our theorem does not rely on the conservation of momentum or on the uncertainty principle.





\section{Our assumptions and theorem}

In this section we present our result, Theorem \ref{theorem1}, stating that spacetime degrees of freedom \edit{violate a Bell inequality} under some assumptions. Thus, \edit{as we argue,} if our assumptions are true then spacetime cannot be classical, \edit{according to a classicality criterion according to which a classical spacetime cannot have spacetime degrees of freedom in an entangled state with a quantum system. We} state and discuss our assumptions below.


\subsection{\edit{A}ssumptions in Bell's theorem}

The following assumptions are standard in Bell's \edit{theorem \cite{Bell}} and other theorems in quantum foundations (e.g., \cite{PBR12,BUGLTCPW20}).

\begin{assumption}[No-superdeterminism]
\label{free}
It is possible to choose \emph{free variables}, i.e. variables with probability distributions that are independent of any events outside their causal \edit{future}.
\end{assumption}

Without this assumption, the events of every experiment could be predetermined. This would not contradict assumption \ref{causality} \edit{below} because in principle there could be a common cause for every experiment, the big bang, for example.

\begin{assumption}[Relativistic causality]
\label{causality}
The outcome probabilities of any experiment \edit{cannot depend on the values of any free variables chosen outside the causal past of the experiment.}
 \end{assumption}

This implies that the outcome probabilities of any experiment cannot depend on \edit{the values of any free variables chosen at spacelike separation of the experiment}, i.e. \emph{no-superluminal signalling}; or on \edit{the values of any free variables chosen in the causal future of the experiment}, i.e. \emph{no-retrocausality}. QT is consistent with this assumption. In particular, \edit{the} no-signalling principle of QT says that measuring distant quantum systems on entangled states does not allow us to transmit any signals.

\begin{assumption}[Background Minkowski spacetime]
\label{minkowski}
The background spacetime near the Earth surface and in the interplanetary space in our solar system is Minkowski. Furthermore, any Bell experiment in the interplanetary space involving quantum systems cannot substantially modified the spacetime geometry. In particular, there cannot be wormholes, closed timelike curves, or any other mechanisms allowing signalling between spacelike separated regions.
\end{assumption}

Making this assumption explicitly might seem excessive. But it is needed to discard the speculations that quantum systems in entangled states are connected via wormholes \cite{MS13} or via other signalling mechanisms (e.g., tachyons \cite{F67}). We note that distant entangled particles allowing superluminal signalling via wormholes would not necessarily contradict relativistic causality (assumption \ref{causality}) because wormholes are possible solutions of Einstein's equations. 

In principle, a rigorous \edit{Bell experiment} in the interplanetary space does not need the background spacetime there to be Minkowski, if wormholes or other superluminal signalling mechanisms are excluded. But, in practice, the experimenters must have a sufficiently good description of the spacetime geometry involved. For this reason, it is a requirement that the background spacetime be Minkowski. Furthermore, imposing this condition on the interplanetary space, at least in an experiment between a laboratory on Earth and a space-based laboratory separated by 
approximately 0.1 light seconds, 
might be needed to close the collapse-locality loophole with quantum systems \cite{K05}. The thought experiment used in the proof of our theorem might require to be implemented even at larger scales in the interplanetary space, or beyond. In the latter case, assumption \ref{minkowski} must be extended so that the background spacetime where the experiment takes place be Minkowski.


\subsection{Assumptions on quantum state reduction}

\begin{assumption}[Superluminal Quantum State Reduction]
\label{assu1}
The quantum state reduction in a quantum measurement is a physical process taking a finite time. Furthermore, in a known reference frame $F_\text{red}$ in a background Minkwoski spacetime, there exists a sufficiently large distance $D_\text{red}$ such that if a pair of quantum systems $A$ and $B$ is prepared in the singlet state and can be kept in this state while they are sent to distant locations separated by a distance $D>D_\text{red}$, then upon measuring $A$ the reduction of the local quantum states of $A$ and $B$ is completed in a time $T_\text{red}=\frac{D_\text{red}}{c}<\frac{D}{c}$, where $c$ is the speed of light through vacuum.
\end{assumption}

Assumption \ref{assu1} is a debatable assumption that seems to contradict our intuitions about relativistic causality. This is because according to relativity theory, there is an inertial reference frame connected to $F_\text{red}$ by a Lorentz transformation in which the quantum state reduction of system $B$ takes place before the quantum measurement is implemented on system $A$. However, this assumption is not inconsistent with relativistic causality because, according to QT, QSR does not allow us to send any superluminal signals. More precisely, QT satisfies the no-signalling principle stating that the outcome probabilities of any quantum measurement on $B$ are independent of the quantum measurement applied on $A$, and vice versa. Moreover, it would not be inconsistent with special relativity that there existed a preferred reference frame in nature with respect to which QSR
propagated instantaneously. Furthermore, as discussed in the introduction, GR does not say anything about quantum superpositions or about quantum entanglement, as it assumes that matter-energy is in well-defined classical values. The truth is that we do not understand QSR at present. 
\edit{Assuming that QSR is a real physical process, the observed violation of Bell inequalities (e.g., \cite{HBDRKBRVSAAPMMTEWTH15,GVMWHHPSKLAAPMBGLSNSUWZ15,SMCBWSGGHACDHLVLTMZSAAPJMKBMKN15}) 
suggests that assumption \ref{assu1} is plausible.}



However, we clarify that there is not consensus about 
the validity of assumption \ref{assu1}. In particular, this assumption was implicitly made by Eppley and Hannah \cite{EH77} in their argument for the necessity of quantizing the gravitational field. Kent \cite{K18} criticised this assumption, suggesting that QSR is not superluminal \cite{K05}. His criticism is based on the logical possibility that a version of
standard QT, with localized QSR obeying the Born rule, can be combined with a theory of gravity that interacts with the local quantum state. Kent supports this criticism with the collapse-locality loophole \cite{K05}, stating that the quantum measurements in previous Bell experiments could have been timelike separated because QSR could take longer than what assumed in the experiments, and in his proposed causal quantum theory \cite{K05,K18.2.1} stating that the violation of Bell inequalities will not be observed in experiments that close this loophole. A Bell experiment \cite{SBHGZ08} has addressed the collapse-locality loophole and claimed to have closed it, under a specific hypothesis for QSR by Diosi \cite{D87} and Penrose \cite{P96}. However, according to Kent, this loophole is still open, and stronger tests have been recently proposed by him to close it in future Bell experiments \cite{K20}.

\subsection{Assumptions on spacetime and quantum theory}


\begin{assumption}[Perfect Distinguishability of Classical Spacetime Geometries]
\label{assu3}
In a background Minkowski spacetime, the location $x$ of a system of mass $m$ as a function of time $t$ can be set in two different well defined classical worldlines $x_0(t)$ and $x_1(t)$, represented by classical states $\lvert x_0(t)\rangle$ and $\lvert x_1(t)\rangle$ 
generating respective well defined classical states $\lvert G_0\rangle$ and $\lvert G_1\rangle$ for spacetime geometries in a spacetime region $R_\text{geom}$.
A `spacetime measurement' experiment $\mathcal{SM}$ can be implemented in $R_\text{geom}$ to perfectly distinguish whether the spacetime geometry there is in the state $\lvert G_0\rangle$ or $\lvert G_1\rangle$. If assumption \ref{assu1} holds, then the spacetime region $R_\text{geom}$ has a time interval $T_\text{geom}>0$ in the reference frame $F_\text{red}$.


\end{assumption}



\edit{We clarify the concept of classical states used above. The worldlines  $x_0 (t)$ and $x_1 (t)$ are two different worldlines according to GR in a background Minkowski spacetime that one could obtain in a classical world that is not governed by quantum theory. In other words, $x_0 (t)$ and $x_1 (t)$ represent worldlines using a set of coordinates for spacetime that are not in any quantum superposition, and in this sense we call them ``classical". The used spacetime coordinates can be transformed according to relativity theory and do not include any sets of coordinates that can be expressed as superpositions of any other sets of valid coordinates. The states $\lvert x_0(t)\rangle$ and $\lvert x_1(t)\rangle$ merely represent the classical worldlines $x_0 (t)$ and $x_1 (t)$. The states $\lvert x_0(t)\rangle$ and $\lvert x_1(t)\rangle$ can be expressed as vector elements of an orthonormal basis in a Hilbert space. In this chosen basis, the vectors for $\lvert x_0(t)\rangle$ and $\lvert x_1(t)\rangle$ are not in superpositions of any vectors in the Hilbert space. This is consistent with our understanding of classical states in quantum theory. 

The states $\lvert G_0\rangle$ and $\lvert G_1\rangle$ represent spacetime geometries generated by the classical worldlines $x_0 (t)$ and $x_1 (t)$ according to GR. The states $\lvert G_0\rangle$ and $\lvert G_1\rangle$ can be expressed as vector elements of an orthonormal basis in a Hilbert space. In this basis, the states $\lvert G_0\rangle$ and $\lvert G_1\rangle$ are not in superpositions of any other vectors in the Hilbert space. In this sense, we say that the states $\lvert G_0\rangle$ and $\lvert G_1\rangle$ are classical.

In other words, in assumption \ref{assu3}, the states $\lvert x_0(t)\rangle$, $\lvert x_1(t)\rangle$, $\lvert G_0\rangle$ and $\lvert G_1\rangle$ merely represent what is predicted by GR in a background Minkowski spacetime.

Assumption \ref{assu3}} is consistent with the predictions of GR. A massive system with well defined classical position $x_i(t)$ as a function of time generates a 
well defined classical state $\lvert G_i\rangle$ for the spacetime geometry in the spacetime region comprising the causal future of $x_i(t)$, according to Einstein's equations. However, Heisenberg's \cite{Heisenbergbook} uncertainty principle of QT states that no physical system can simultaneously have perfectly well defined position and momentum at any given time. But if the system's mass $m$ is large enough then the uncertainties $\Delta x$ and $\Delta v$ for its potion and velocity can be negligible, while satisfying $m\Delta v\Delta x \gtrapprox h$, according to the uncertainty principle, where $h$ is Planck's constant. 

Furthermore, in this case, if the world lines $x_0(t)$ and $x_1(t)$ are chosen appropriately, then the states $\lvert G_0\rangle$ and $\lvert G_1\rangle$ can in principle be distinguished in an experiment. For example, a source system of mass $m_\text{S}=m$ may be set in one of two free falling trajectories $x_0(t)$ and $x_1(t)$ separated by a distance $d$ during a time $T_\text{geom}$, and the states $\lvert G_0\rangle$ and $\lvert G_1\rangle$ could be distinguished by measuring the final position of a probe system with mass $m_\text{P}$ free falling simultaneously with the source system and separated by some distance $d'$ from both possible trajectories $x_0(t)$ and $x_1(t)$ of the source system, with appropriately chosen values of $m_\text{S},m_{\text{P}},d,d'$ and $T_\text{geom}$. 
This experiment can be described classically in the Newtonian limit of GR within very good approximation, if the uncertainties $\Delta x_\text{P}$ ($\Delta x_\text{S}$) and $\Delta v_\text{P}$ ($\Delta v_\text{S}$) for the position and velocity of the massive probe (source) system are guaranteed to be small enough during the experiment. In particular, we may choose parameters consistent with the uncertainty principle for the source and probe systems, satisfying $m_\text{A}\Delta x_\text{A}<<Gm_\text{S}m_\text{P}T^2_{\text{geom}}/{d'^2}<m_\text{A}d'< m_\text{A}d$ and $m_\text{A}\Delta v_{\text{A}} T_\text{geom}<<Gm_\text{S}m_\text{P}T^2_{\text{geom}}/{d'^2}$, for $``\text{A}"=``\text{P}",``\text{S}"$, where $G$ is the gravitational constant. This experiment can in principle be implemented in space with a low background gravitational field so that the source and probe masses can free fall for as long as it is required.

\begin{assumption}[Macroscopic Quantum Superpositions]
\label{assu5}
The location of a physical system of mass $m$ as a function of time can be set in an arbitrary quantum superposition $\lvert\Psi(t)\rangle=C_0\lvert x_0(t)\rangle+C_1\lvert x_1(t)\rangle$ of the classical worldlines $x_0(t)$ and $x_1(t)$ satisfying assumption \ref{assu3}\edit{, where $C_0,C_1\in\mathbb{C}$ and $\lvert C_0\rvert^2+\lvert C_1\rvert^2=1$}. 
Furthermore, this quantum superposition can be sustained during a time interval $T_\text{mass}>T_\text{geom}$ in the reference frame $F_\text{red}$ if assumption \ref{assu1} holds.
\end{assumption}

A crucial observation is that the parameters $m$ and $T_\text{geom}$ of assumption \ref{assu3} might need to be large enough in order to be able to distinguish the classical states $\lvert G_0\rangle$ and $\lvert G_1\rangle$ in an experiment that it might not be possible to prepare and maintain a system of mass $m$ in an arbitrary quantum superposition $\lvert\Psi(t)\rangle=C_0\lvert x_0(t)\rangle+C_1\lvert x_1(t)\rangle$ during a time interval $T_\text{mass}>T_\text{geom}$. This could be due to some fundamental decoherence effects, for example, due to gravitational effects \cite{BGU17}. \edit{This could also be due to non-linear effects (e.g. \cite{C08}).}
Thus, it is conceivable that assumption \ref{assu5} does not hold, establishing a limit of QT.

\begin{assumption}[Spacetime Quantum State Reduction]
\label{assu6}
If assumption \ref{assu1}, \ref{assu3} and \ref{assu5} hold, then the spacetime measurement $\mathcal{SM}$ either i) does not change the quantum state $\lvert \Psi(t)\rangle$; or ii) reduces it to $\lvert x_i(t)\rangle$ and outputs $s=i$ with probability $\lvert C_i\rvert^2$, within a time interval $T_\text{geom}$ in the reference frame $F_{red}$, for all $i\in\{0,1\}$.
\end{assumption}

Certainly, $\mathcal{SM}$ could modify the quantum state $\lvert \Psi(t)\rangle$ in ways not considered by assumption \ref{assu6}. Transformations that lie outside our current understanding of QT are deliberately neglected, as one of our aims here is to investigate possible limits on QT that could arise if spacetime were fundamentally classical. However, there are transformations consistent with QT that $\mathcal{SM}$ could induce that are not considered in assumption \ref{assu6}.

Importantly, we have neglected the possibility that $\mathcal{SM}$ comprises only unitary evolution of all physical systems involved. This could be consistent with the ``many worlds" interpretation of QT. But this would contradict our assumptions here that QSR is a real physical process. Moreover, if $\mathcal{SM}$ only implements a unitary evolution, then it follows that some spacetime degrees of freedom become entangled with other systems, implying also that spacetime is non-classical, according to a natural criterion for non-classicality.

Nevertheless, there are other transformations consistent with QT that could be induced by $\mathcal{SM}$. For example, $\mathcal{SM}$ could apply a unitary operation, changing the amplitude $C_i$ to $C_i'$, with $\lvert C_i'\rvert^2\neq\lvert  C_i\rvert^2$, and then reduce the quantum state to $\lvert x_i(t)\rangle$ and output $s=i$ with probability $\lvert C_i'\rvert^2$, for all $i\in\{0,1\}$. We do not aim here to consider all possible transformations consistent with QT that $\mathcal{SM}$ can implement. We focus on the two possibilities stated in assumption \ref{assu6}, which appear to us as the simplest and most natural to explore. 

Our assumption that a measurement of spacetime degrees of freedom may induce a reduction of the quantum state is motivated by the hypothesis that gravity, and hence spacetime, plays a fundamental role in QSR \cite{K66,D89,GGR90,P96}


In addition to assumption \ref{assu5}, the following two assumptions only concern the limits of validity of QT.

\begin{assumption}[Experimentally Verified Quantum Theory]
\label{assu4}
Quantum theory is valid in the microscopic scales and in the situations in which it has been experimentally verified.
\end{assumption}

This assumption seems trivially obvious. But we state it explicitly in order to emphasize that 
assumption \ref{assu5} and the following assumption, which have not been confirmed in experiment, might not hold.

\begin{assumption}[Long Range Quantum Entanglement]
\label{assu7}
In a background Minkowski spacetime, two quantum systems $A$ and $B$ prepared in the singlet state can be kept in this state while they are sent to distant locations separated by a sufficiently large distance $D_\text{ent}$ such that if assumptions \ref{assu1} and \ref{assu3} hold then $D_\text{ent}>c\bigl(T_\text{red}+T_\text{geom}+T_\text{extra}\bigr)$, in the reference frame $F_\text{red}$, for some sufficiently long time interval $T_\text{extra}>0$.
\end{assumption}

Recently, quantum entangled photons have been distributed between two cities in China separated by 1,200 km via the Micius satellite \cite{Yinetal17}. However, satisfying assumption \ref{assu7} might require entanglement distribution between interplanetary distances, or beyond. 
According to QT, this is in principle possible. In particular, since the interplanetary space has much less atoms 
than the Earth atmosphere, it is in principle easier for a photon to travel a given distance in the interplanetary space than to travel that distance in the Earth atmosphere. Furthermore, if distributing a pair of entangled photons through a large distance $D_\text{ent}$ succeeds with a small probability $P_\text{ent}>0$, by trying with $N>>(P_\text{ent})^{-1}$ pairs, at least one pair of entangled photons will be distributed successfully with high probability. 
Nevertheless, it is conceivable that there is an unknown limit on QT stating that assumption \ref{assu7} cannot hold. As for assumption \ref{assu5}, this could be due to fundamental decoherence effects, for example, due to \edit{gravity \cite{BGU17}}.

\begin{theorem}
\label{theorem1}
If assumptions \ref{free} -- \ref{assu7} hold then there \edit{is a Bell experiment with a quantum system $A$ at Alice's wing and spacetime degrees of freedom $S$ at Bob's wing with respective binary inputs $\alpha,\beta$ and outputs $a,s$, producing a probability distribution $P_\text{AS}(as\vert \alpha\beta)$ that violates the CHSH inequality.} 
\end{theorem}

\edit{Below we} prove the theorem by considering a thought experiment, making assumptions \ref{free} -- \ref{assu7}. \edit{We argue that if assumptions \ref{free} -- \ref{assu7} hold then Theorem \ref{theorem1} implies that spacetime cannot be considered classical, and in this sense has quantum features. We make the following assumption.

\begin{assumption}[A classicality criterion]
\label{classicality}
If spacetime can be reasonably considered classical then any spacetime degrees of freedoom can be described by a classical state subject to classical measurements within the quantum formalism, and in particular, cannot be entangled with any quantum system.
\end{assumption}


Our argument works by contradiction. We assume that $S$ is a classical system. From assumptions \ref{free} and \ref{classicality}, the joint system $AS$ can be described by a non-entangled quantum state $\rho_{AS}$, which is given by
\begin{equation}
\label{classicalstate}
\rho_{AS}=\int_\Lambda d\lambda P(\lambda)\bigl(\rho_A^\lambda\otimes \rho_S^\lambda\bigr),
\end{equation}
where $\rho_A^\lambda$ are density operators for the possible quantum states of $A$ and $\rho_S^\lambda$ are density operators for the possible classical states of $S$, and where $\Lambda$ is an arbitrary set of variables generated within the union of the causal pasts of the -- spacelike separated -- spacetime regions where the free variables $\alpha$ and $\beta$ were chosen by Alice and Bob. We note in particular that the probability distribution $\{P(\lambda)\}_{\lambda\in\Lambda}$ does not depend on the free variables $\alpha$ and $\beta$, as follows from assumption \ref{free}. Thus, the probability distribution $P_\text{AS}(as\vert \alpha\beta)$ is given by
\begin{eqnarray}
\label{classicaldist}
P_\text{AS}(as\vert \alpha\beta)&=&\int_\Lambda d\lambda P(\lambda)\text{Tr}\bigl[\bigl(\rho_A^\lambda\otimes \rho_S^\lambda\bigr)\bigl(M_A^{a,\lambda}\otimes M_S^{s,\lambda}\bigr)\bigr]\nonumber\\
&=&\int_\Lambda d\lambda P(\lambda)P_\text{A}(a\vert \alpha,\lambda)P_\text{S}(s\vert \beta,\lambda),
\end{eqnarray}
where $M_A^{a,\lambda}$ and $M_S^{s,\lambda}$ are quantum and classical measurement operators acting on the systems $A$ and $S$, respectively. From (\ref{classicaldist}), the probability distribution $P_\text{AS}(as\vert \alpha\beta)$ cannot violate a Bell inequality \cite{BCPSW14}. This contradicts Theorem \ref{theorem1}. It follows that spacetime is not classical according to the criterion of assumption \ref{classicality}.

}

\section{A thought experiment and proof of Theorem \ref{theorem1}}

\begin{figure*}
\includegraphics[scale=0.3]{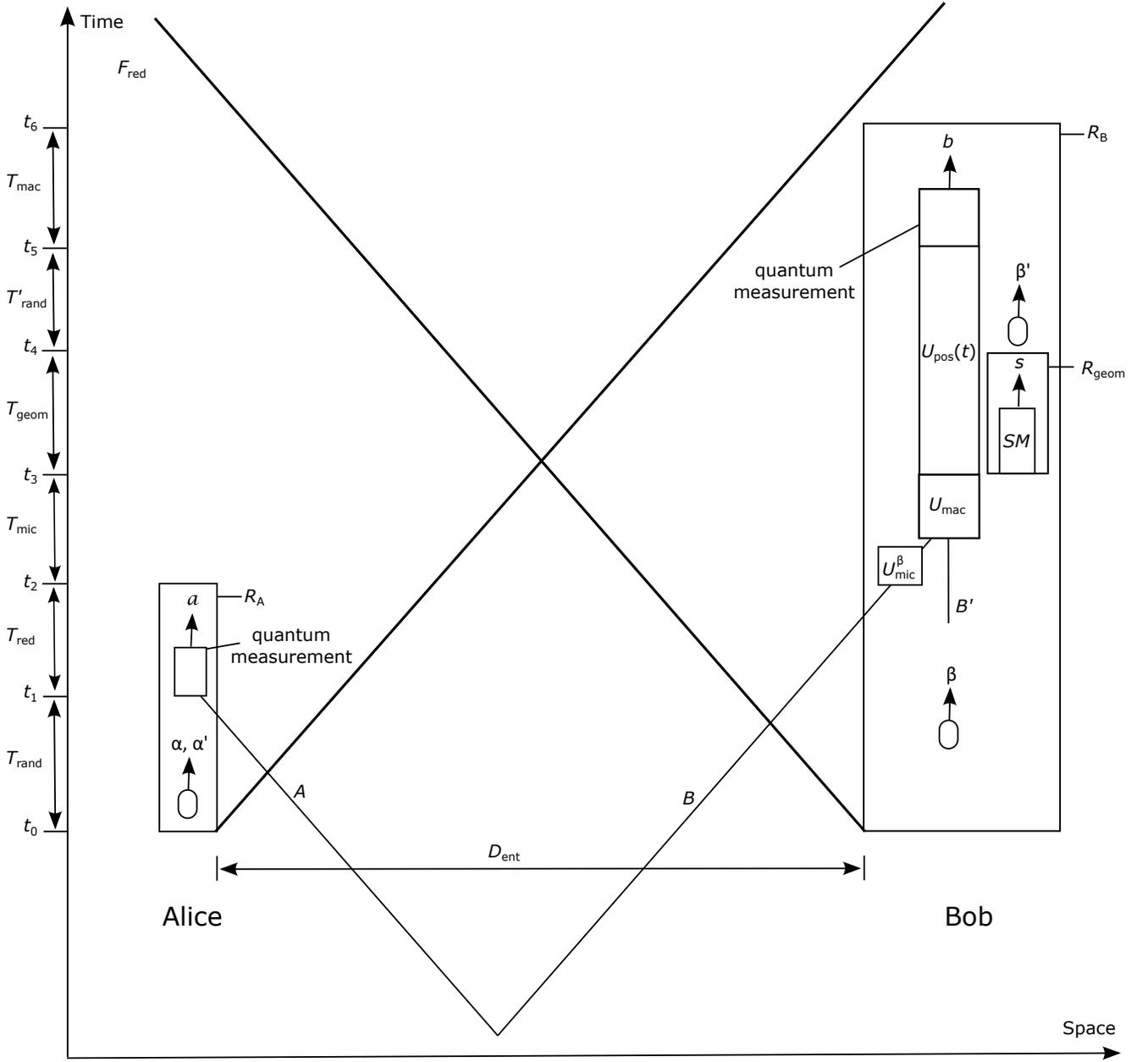}
\caption{\label{figure} \textbf{Our thought experiment.} We illustrate the thought experiment used to show Theorem \ref{theorem1}, in the reference frame $F_\text{red}$ in a background Minkowski spacetime, as described in the main text. Alice's and Bob's actions take place within the spacetime regions $R_\text{A}$ and $R_\text{B}$, respectively, which are spacelike separated, as illustrated by the light rays (thick diagonal lines). The generation of random numbers \edit{(i.e., free variables)} in the experiment is illustrated by the rounded rectangles. The short vertical arrows denote classical outputs.}
\end{figure*}

Alice and Bob perform the following experiment near the Earth surface, or in the interplanetary space in our solar system. Thus, from assumption \ref{minkowski}, the experiment takes place in a background Minkowski spacetime. According to assumption \ref{assu1}, the reference frame $F_\text{red}$ is known. We can then assume that Alice and Bob know this reference frame. We describe the experiment in this frame (see Fig. \ref{figure}).

Let $T_\text{rand},T_\text{rand}',T_{\text{mic}},T_{\text{mac}}>0$ be time intervals defined below, and let
\begin{equation}
\label{f1}
T_\text{extra}=T_\text{rand}+T_\text{rand}'+T_{\text{mic}}+T_{\text{mac}}.
\end{equation}
Alice and Bob are separated by a distance 
\begin{equation}
\label{f2}
D_\text{ent}>c(T_\text{red}+T_\text{geom}+T_\text{extra}).
\end{equation}
Alice and Bob receive microscopic quantum systems $A$ and $B$ encoding qubits in the singlet state (\ref{e0}).
This is possible according to assumption \ref{assu7}.


Between times $t_0=0$ and $t_1=T_\text{ran}$, Alice generates two random bits $\alpha$ and $\alpha'$\edit{, which we assume are free variables (assumption \ref{free})}. Then, at time $t_1$, she receives the qubit $A$ and measures it immediately upon reception in the qubit orthonormal basis $\mathcal{B}_{\alpha\alpha'}=\{\lvert \phi_{\alpha\alpha'}^a\rangle\}_{a=0}^1$ and obtains the bit outcome $a$, where 
\begin{eqnarray}
\label{e1}
\lvert \phi_{00}^a\rangle&=&\lvert a\rangle,\nonumber\\
\lvert \phi_{10}^a\rangle&=&\frac{1}{\sqrt{2}}\bigl(\lvert 0\rangle+(-1)^a\lvert 1\rangle\bigr),\nonumber\\
\lvert \phi_{01}^{a}\rangle&=&\cos\Bigl(\frac{\pi}{8}\Bigr)\lvert a\rangle+(-1)^a\sin\Bigl(\frac{\pi}{8}\Bigr)\lvert \overline{a}\rangle,\nonumber\\
\lvert \phi_{11}^{a}\rangle&=&\sin\Bigl(\frac{\pi}{8}\Bigr)\lvert a\rangle+(-1)^a\cos\Bigl(\frac{\pi}{8}\Bigr)\lvert \overline{a}\rangle,
\end{eqnarray}
for all $a\in\{0,1\}$. An important property of the singlet is that it can be expressed by (\ref{e0}) in any qubit orthonormal basis $\{\lvert 0\rangle,\lvert 1\rangle\}$. In particular, we have
\begin{equation}
\label{e1.1}
\lvert\Psi^-\rangle=\frac{1}{\sqrt{2}}\bigl(\lvert \phi_{\alpha\alpha'}^0\rangle\lvert \phi_{\alpha\alpha'}^1\rangle-\lvert \phi_{\alpha\alpha'}^1\rangle\lvert\phi_{\alpha\alpha'}^0\rangle\bigr),
\end{equation}
for all $\alpha,\alpha'\in\{0,1\}$. Thus, Alice obtains outcome $a$ with probability $\frac{1}{2}$, as predicted by QT, for all $a\in\{0,1\}$. This follows from assumption \ref{assu4} because $A$ is a microscopic quantum system. From assumption \ref{assu1}, due to Alice's quantum measurement on $A$, the local quantum states of $A$ and $B$ have reduced, and Alice has obtained the outcome $a$, by the time $t_2=t_1+T_\text{red}>0$. From (\ref{e1.1}), the reduced quantum states of Alice and Bob are $\lvert \phi_{\alpha\alpha'}^a\rangle$ and $\lvert \phi_{\alpha\alpha'}^{\bar{a}}\rangle$, respectively.


Between times $t_0$ and $t_2$, Bob generates a random bit $\beta$\edit{, which we assume is a free variable (assumption \ref{free})}. Then, at time $t_2$ Bob receives the qubit $B$ in the reduced quantum state $\lvert \phi_{\alpha\alpha'}^{\bar{a}}\rangle$, as discussed above. At time $t_2$, immediately upon reception, Bob applies a unitary operation $U_{\text{mic}}^\beta$ on $B$, where $U_{\text{mic}}^0=\mathbb{1}$ is the identity operation and $U_{\text{mic}}^1=U_\text{mic}$ is a unitary operation satisfying 
\begin{eqnarray}
\label{e2}
U_\text{mic}\lvert\phi_{10}^a\rangle &=& (-1)^a\lvert\phi_{00}^a\rangle,\nonumber\\
U_\text{mic}\lvert\phi_{00}^a\rangle &=& \lvert\phi_{10}^{\overline{a}}\rangle,
\end{eqnarray}
for all $a\in\{0,1\}$. This is possible according to assumption \ref{assu4}, because $B$ is a microscopic quantum system. The label `mic' stands for `microscopic'.

After the unitary operation $U_\text{mic}^\beta$ is completed, Bob applies a unitary operation $U_\text{mac}$ on the joint system $BB'$, where $B'$ is a macroscopic  
massive system initially prepared by Bob in a classical 
position state $\lvert x_\text{init}\rangle$. 
The label `mac' stands for `macroscopic'.
Let $m_\text{mic}$  and $m_\text{mac}$ be the respective masses of the system $B$ and $B'$, with $m_\text{mic}<<m_\text{mac}$ and with the mass of the joint system $BB'$ be $m=m_\text{mic}+m_\text{mac}$. In the case that $B$ is a photon, we can simply set $m_\text{mic}=0$. Let Bob's unitary operation $U_\text{mac}$ be completed at the time $t_3=t_2+T_\text{mic}$, and satisfy 
\begin{equation}
\label{e3}
U_\text{mac}\lvert a\rangle\lvert x_\text{init}\rangle=\lvert x_a(t_3)\rangle,
\end{equation}
where $\lvert x_a(t_3)\rangle$ is a classical state for the location $x$ of the joint system $BB'$ at the time $t_3$, corresponding to the worldline $x_a(t)$ satisfying assumptions \ref{assu3} and \ref{assu5}, for all $a\in\{0,1\}$. 
Thus, according to assumption \ref{assu3}, the state $\lvert x_a(t)\rangle$ generates a classical state $\lvert G_a\rangle$  for spacetime geometry in a spacetime region $R_\text{geom}$, 
for all $a\in\{0,1\}$. \edit{We note that we are not assuming that the spacetime geometry is set in a superposition state.} From assumptions \ref{assu1} and \ref{assu3}, $R_\text{geom}$ has a time interval $T_\text{geom}>0$ in the frame $F_\text{red}$. Let $R_\text{geom}$ have time coordinates  $t\in[t_3,t_4]$ in the frame $F_\text{red}$, where $t_4=t_3+T_\text{geom}$.

\edit{We note that (\ref{e3}) is justified by assumption \ref{assu5}. If Bob's system $B$ is in a superposition of states $\lvert 0\rangle$ and  $\lvert 1 \rangle$, as given by one of the states (\ref{e1}), for instance, then it follows from (\ref{e3}) that the location of the system $BB'$ can be set in a superposition. This is consistent with assumption \ref{assu5}.}

Between the times $t_3$ and $t_5$, the position of the system $BB'$ is set to evolve with the unitary operation $U_\text{pos}(t)$ as a function of time $t$, which satisfies
\begin{equation}
\label{position}
U_\text{pos}(t)\lvert x_a(t_3)\rangle=\lvert x_a(t)\rangle,
\end{equation}
for all $t\in[t_3,t_5]$ and all $a\in\{0,1\}$.

From the time $t_3$, Bob performs an experiment $\mathcal{SM}$ in the spacetime region $R_\text{geom}$ to measure whether the spacetime geometry in $R_\text{geom}$ is in the classical state $\lvert G_0\rangle$ or $\lvert G_1\rangle$. Let $S$ denote the degrees of freedom corresponding to the spacetime geometry in the spacetime region $R_\text{geom}$. According to assumption \ref{assu3}, if $S$ is in the state $\lvert G_i\rangle$ then the experiment  $\mathcal{SM}$ outputs the bit $s=i$ with certainty, for all $i\in\{0,1\}$. Bob designs the experiment $\mathcal{SM}$ so that it outputs a bit $s$ with certainty at the time $t_4$. 
For example, if $\mathcal{SM}$ cannot distinguish whether $S$ is in one of the classical states $\lvert G_0\rangle$ or $\lvert G_1\rangle$, it outputs a random bit $s$.

Between the times $t_4$ and $t_5=t_4+T_\text{rand}'$, Bob generates a random bit $\beta'$, \edit{which we assume is a free variable (assumption \ref{free})}, where $T_\text{rand}'>0$. If $\beta'=0$, at the time $t_5$, Bob measures $BB'$ in the basis $\mathcal{B}_\text{mac}=\{\lvert x_b(t_5)\rangle\}_{b=0}^1$ and obtains a bit outcome $b$. The measurement is completed and Bob obtains his outcome by the time $t_6=t_5+T_\text{mac}$, for some $T_\text{mac}>0$.

If $\beta'=1$, at the time $t_5$, Bob applies a unitary operation $U_\text{mac}'$ on $BB'$ satisfying
\begin{equation}
\label{inverse}
U_\text{mac}'\lvert x_a(t_5)\rangle=\lvert a\rangle\lvert x_\text{fin}\rangle,
\end{equation}
where $\lvert x_\text{fin}\rangle$ is a classical state for the position of the system $B'$, for all $a\in\{0,1\}$. The unitary operation $U_\text{mac}'$ is completed by the time $t_6$. In this case, Bob sets $b=0$ with unit probability, by the time $t_6$.


Let $R_\text{A}$ be the spacetime region corresponding to Alice's location with the time coordinates $t\in[t_0,t_2]$ in the reference frame $F_\text{red}$. Let $R_\text{B}$ be the spacetime region corresponding to Bob's location with the time coordinates $t\in[t_0,t_6]$ in the reference frame $F_\text{red}$, satisfying $R_\text{B}\supset R_\text{geom}$. We note that Alice obtains her bits $\alpha,\alpha'$ and $a$ in $R_\text{A}$, and Bob obtains his bits $\beta,\beta',s$ and $b$
 in $R_\text{B}$. It follows from (\ref{f1}) and (\ref{f2}), and from the description of the experiment, that $R_\text{A}$ and $R_\text{B}$ are spacelike separated. Thus, in particular, Bob obtains his bits $\beta,\beta',s$ and $b$ at spacelike separation from Alice obtaining her bits $\alpha,\alpha'$ and $a$.


Let $P(abs\vert \alpha\alpha'\beta\beta')$ be the probability distribution for Alice's and Bob's outcomes $a,b,s$, given their inputs $\alpha,\alpha',\beta,\beta'$, for all $a,b,s,\alpha,\alpha',\beta,\beta'\in\{0,1\}$. We define
\begin{eqnarray}
\label{a1}
P_\text{AB}(ab\vert\alpha\alpha'\beta\beta')&=&\sum_{s=0}^1 P(abs\vert\alpha\alpha'\beta\beta'),\nonumber\\
P_\text{AS}(as\vert\alpha\alpha'\beta\beta')&=&\sum_{b=0}^1 P(abs\vert\alpha\alpha'\beta\beta'),\nonumber\\
P_\text{BS}(bs\vert\alpha\alpha'\beta\beta')&=&\sum_{a=0}^1 P(abs\vert\alpha\alpha'\beta\beta'),
\end{eqnarray}
for all $a,b,s,\alpha,\alpha',\beta,\beta'\in\{0,1\}$.

We show below that 
\begin{equation}
\label{a5.2}
P_\text{AS}(as\vert \alpha\alpha'\beta 0)=\frac{1}{2}\bigl\lvert \langle \phi_{\alpha\alpha'}^{a}\vert \phi_{\beta0}^{\overline{s}}\rangle\bigr\rvert^2,
\end{equation}
for all 
$a,s,\alpha,\alpha',\beta\in\{0,1\}$. 
From assumption \ref{causality}, the probability distribution for the outcomes $a$ and $s$ cannot depend on the value of $\beta'$ because this is a free variable (assumption \ref{free}) chosen at spacelike separation from Alice obtaining $a$ (no-superluminal signalling) and in the causal future of Bob obtaining $s$ (no-retrocausality). Thus, it follows from (\ref{a5.2}) that
\begin{equation}
\label{newa5.2}
P_\text{AS}(as\vert \alpha\alpha'\beta \beta')=\frac{1}{2}\bigl\lvert \langle \phi_{\alpha\alpha'}^{a}\vert \phi_{\beta0}^{\overline{s}}\rangle\bigr\rvert^2,
\end{equation}
for all $a,s,\alpha,\alpha',\beta,\beta'\in\{0,1\}$. 

Crucially, the probability distribution $P_\text{AS}(as\vert \alpha 1 \beta 1)$ given by (\ref{newa5.2}), i.e., when Alice sets $\alpha'=1$ and Bob sets $\beta'=1$, violates the CHSH inequality (\ref{chsh}) and saturates the Tsirelson bound (\ref{cirel}), by taking $s$ as Bob's output (see Appendix \ref{appchsh}). In the case $\beta'=1$, Bob does not apply any quantum measurement on his system $BB'$. Bob's only measurement in this case is the spacetime measurement $\mathcal{SM}$ of the spacetime degrees of freedom $S$, with outcome $s$. 
\edit{This proves Theorem \ref{theorem1}.}

\subsection{Proof of (\ref{a5.2})}

In what follows we consider the case $\beta'=0$ and show (\ref{a5.2}). An important part of our proof comprises to show that
\begin{equation}
\label{b1}
P(ab\overline{b}\vert\alpha\alpha'\beta 0)=0,
\end{equation}
for all $a,b,\alpha,\alpha',\beta\in\{0,1\}$. To show (\ref{b1}), we consider first the case $\alpha'=0$ and $\beta=\alpha$. In this case, Alice measures her qubit $A$ in the basis $\mathcal{B}_{\alpha0}=\{\lvert \phi_{\alpha0}^{a}\rangle\}_{a=0}^1$ and obtains a bit outcome $a$ with probability $\frac{1}{2}$.
Thus, from (\ref{e1.1}), Alice's qubit state reduces to $\lvert \phi_{\alpha0}^{a}\rangle$ and Bob's qubit state reduces to $\lvert \phi_{\alpha0}^{\bar{a}}\rangle$. 
Then it is straightforward to obtain from (\ref{e2}) that after Bob applies the unitary operation $U_{\text{mic}}^\beta$, his qubit $B$ transforms into the state $\lvert \overline{a}\rangle$, up to a global phase. Thus, from (\ref{e3}), after applying the unitary operation $U_\text{mac}$, the position state of Bob's joint quantum system $BB'$ transforms into the classical state $\lvert x_{\overline{a}}(t_3)\rangle$ at time $t_3$ (up to a global phase). From (\ref{position}), the worldline of $BB'$ is given by the classical state $\lvert x_{\overline{a}}(t)\rangle$, in the time interval $t\in[t_3,t_5]$ (up to a global phase). From assumption \ref{assu3}, this state generates the classical spacetime geometry state $\lvert G_{\overline{a}}\rangle$ (up to a global phase) for the spacetime degrees of freedom $S$ in the spacetime region $R_\text{geom}$. Thus, Bob's spacetime measurement experiment $\mathcal{SM}$ outputs $s=\overline{a}$ in $R_\text{geom}$with certainty, and by the time $t_4$. From assumption \ref{assu6}, the experiment $\mathcal{SM}$ does not change the quantum state of the joint system $BB'$; or it reduces it according to the Born rule, in this case to its previous state $\lvert x_{\overline{a}}(t_4)\rangle$  
at the time $t_4=t_3+T_\text{geom}$ with unit probability. Therefore, Bob's joint system $BB'$ remains in the quantum state $\lvert x_{\overline{a}}(t_4)\rangle$ after the experiment $\mathcal{SM}$, at time $t_4$. Then, from (\ref{position}), the location of $BB'$ continues to evolve in the worldline given by the classical state $\lvert x_{\overline{a}}(t)\rangle$, during the time interval $t\in[t_4,t_5]$. Thus, at the time $t_5$, the location of $BB'$ is given by the state $\lvert x_{\overline{a}}(t_5)\rangle$. Therefore, Bob's quantum measurement in the basis $\mathcal{B}_\text{mac}$ at the time $t_5$ gives the outcome $b=\overline{a}$ by the time $t_6$ with certainty. Thus, it follows from the observation above that Alice obtains the outcome $a$ with probability $\frac{1}{2}$, that
\begin{equation}
\label{a2}
P(a\overline{a}\overline{a}\vert \alpha0\alpha0)=\frac{1}{2},
\end{equation}
for all $a,\alpha\in\{0,1\}$. It follows from (\ref{a2}) that
\begin{equation}
\label{a4}
P(aaa\vert \alpha0\alpha0)=0,
\end{equation}
for all $a,\alpha\in\{0,1\}$. Thus, we have
\begin{eqnarray}
\label{a5}
P_\text{BS}(bb\vert \alpha0\alpha 0)&=&\sum_{a=0}^1P(abb\vert \alpha0\alpha 0)\nonumber\\
&=&P(\overline{b}bb\vert \alpha0\alpha 0)\nonumber\\
&=&\frac{1}{2},
\end{eqnarray}
for all $b,\alpha\in\{0,1\}$, where in the first line we used (\ref{a1}), in the second line we used (\ref{a4}), and in the third line we used (\ref{a2}). 

Then, since Bob obtains his outcomes $b$ and $s$ at spacelike separation from Alice generating her inputs $\alpha$ and $\alpha'$, the principle of no-superluminal signalling (assumption \ref{causality}) requires that Bob's probability distribution $P_\text{BS}(bs\vert\alpha\alpha'\beta 0)$ be independent of the values of $\alpha$ and $\alpha'$, otherwise Alice could send a superluminal signal to Bob by choosing appropriate values of $\alpha$ and $\alpha'$. Thus, from (\ref{a5}), we obtain
\begin{equation}
\label{a3}
P_\text{BS}(bb\vert \alpha \alpha'\beta 0)=\frac{1}{2},
\end{equation}
for all $b,\alpha,\alpha',\beta\in\{0,1\}$. It follows from (\ref{a3}) that
\begin{equation}
\label{c1}
P_{\text{BS}}(b\overline{b}\vert\alpha\alpha'\beta 0)=0,
\end{equation}
for all $b,\alpha,\alpha',\beta\in\{0,1\}$. Then, from (\ref{a1}) and (\ref{c1}), we have
\begin{equation}
\label{c2}
\sum_{a=0}^1P(ab\overline{b}\vert\alpha\alpha'\beta 0)=0,
\end{equation}
for all $b,\alpha,\alpha',\beta\in\{0,1\}$. Thus, (\ref{b1}) follows from (\ref{c2}).

Our proof of (\ref{a5.2}) considers two broad cases, in agreement with assumption \ref{assu6}. In the first case, we assume that Bob's measurement of the spacetime degrees of freedom $S$ in the experiment $\mathcal{SM}$ does not change the quantum state of his system $BB'$. In the second case, we assume that Bob's experiment $\mathcal{SM}$ reduces the quantum state of $BB'$ within the time interval $T_\text{geom}$ according to the Born rule. 

We show (\ref{a5.2}) in the case that $\mathcal{SM}$ does not change the quantum state of $BB'$. In this case, we can determine the probability distribution $P_\text{AB}(ab\vert \alpha\alpha'\beta 0)$ by ignoring the action of $\mathcal{SM}$ and assuming that Bob's actions are only the unitary operations $U_\text{mic}^\beta$, $U_\text{mac}$ and $U_\text{pos}(t)$, followed by the quantum measurement $\mathcal{B}_{\text{mac}}$. This situation is described purely by quantum theory. 
Thus, in this case, it is straightforward to show (see Appendix \ref{app2}) that
\begin{equation}
\label{b2}
P_\text{AB}(ab\vert \alpha\alpha'\beta 0)=\frac{1}{2}\bigl\lvert \langle \phi_{\alpha\alpha'}^{a}\vert \phi_{\beta0}^{\overline{b}}\rangle\bigr\rvert^2,
\end{equation}
for all $a,b,\alpha,\alpha',\beta\in\{0,1\}$. Thus, we have
\begin{eqnarray}
\label{b3}
P_\text{AS}(ab\vert \alpha\alpha'\beta0)&=&\sum_{b'=0}^1P(ab'b\vert \alpha\alpha'\beta 0)\nonumber\\
&=&P(abb\vert \alpha\alpha'\beta 0)\nonumber\\
&=&\sum_{s=0}^1P(abs\vert \alpha\alpha'\beta 0)\nonumber\\
&=&P_\text{AB}(ab\vert \alpha\alpha'\beta 0)\nonumber\\
&=&\frac{1}{2}\bigl\lvert \langle \phi_{\alpha\alpha'}^{a}\vert \phi_{\beta0}^{\overline{b}}\rangle\bigr\rvert^2,
\end{eqnarray}
for $a,b,\alpha,\alpha',\beta\in\{0,1\}$, as claimed; where in the first line we used (\ref{a1}); in the second and third lines we used (\ref{b1}); in the fourth line we used (\ref{a1}); and in the last line we used (\ref{b2}).

Now we show (\ref{a5.2}) in the case that $\mathcal{SM}$ reduces the quantum state of $BB'$ within the time interval $T_\text{geom}$ according to the Born rule, as stated by assumption \ref{assu6}. At the time $t_3$, Bob's joint system $BB'$ has location given by the quantum state
\begin{equation}
\label{x1}
\lvert \Psi(t_3)\rangle=C_0\lvert x_0(t_3)\rangle+C_1\lvert x_1(t_3)\rangle,
\end{equation}
for some $C_0,C_1\in\mathbb{C}$ satisfying $\lvert C_0\rvert^2+\lvert C_1\rvert^2=1$.
If left undisturbed, this quantum state evolves with the unitary operation $U_\text{pos}(t)$ satisfying (\ref{position}), between the time interval $t\in[t_3,t_5]$. Thus, from (\ref{x1}), and from assumptions \ref{assu3}, \ref{assu5} and \ref{assu6}, $\mathcal{SM}$ reduces the quantum state of $BB'$ to $\lvert x_i(t_4)\rangle$ and outputs $s=i$ with probability $\lvert C_i\rvert^2$ at the time $t_4$, 
for all $i\in\{0,1\}$.  Then, from (\ref{position}), this state evolves unitarily to $\lvert x_i(t_5)\rangle$ at the time $t_5$. This means that the joint action of the unitary operation $U_\text{pos}(t)$ from $t_3$ to $t_5$ and the spacetime measurement $\mathcal{SM}$ from $t_3$ to $t_4$ 
modifies the quantum state for the location of $BB'$ in the same way as the joint action of the unitary operation $U_\text{pos}(t)$ from $t_3$ to $t_5$ and the quantum measurement $\mathcal{B}_\text{mac}=\{\lvert x_0(t_5)\rangle,\lvert x_1(t_5)\rangle\}$ at the time $t_5$ with outcome $s\in\{0,1\}$. Thus, in this case, the probability distribution $P_{AS}(as\vert \alpha\alpha'\beta 0)$ can be derived by assuming that Bob's actions are the unitary operation $U_\text{mic}^\beta$ on $B$, followed by the unitary operations $U_\text{mac}$ and $U_\text{pos}(t)$ with $t\in[t_3,t_5]$ on $BB'$, and then by the quantum measurement $\mathcal{B}_\text{mac}$ on $BB'$ with outcome $s\in\{0,1\}$. The probability distribution $P_{AB}(ab\vert \alpha\alpha'\beta 0)$ given by (\ref{b2}) was derived precisely under these assumptions. Thus, in this case, $P_{AS}(as\vert \alpha\alpha'\beta 0)$ is equivalent to the probability distribution $P_{AB}(ab\vert \alpha\alpha'\beta 0)$ given by (\ref{b2}), by replacing $B$ by $S$ and $b$ by $s$, for all $a,b,s,\alpha,\alpha',\beta\in\{0,1\}$. Therefore, (\ref{a5.2}) follows from (\ref{b2}).

\section{Discussion}

Arguments for the necessity of quantizing gravity and spacetime can shed light about limits that QT must have if gravity and spacetime are fundamentally classical. For example, Kent’s \cite{K18} refutation of the Eppley-Hannah \cite{EH77} argument is based on the logical possibility that a modification of
standard QT, with localized quantum state reduction obeying the Born rule, can be combined
with a theory of gravity that interacts with the local quantum state. Other refutations \cite{HC01,MOS20} of the Eppley-Hannah argument reason that momentum might not be conserved fundamentally. Moreover, a recent argument \cite{MV17.2} has been evaded by the construction of a theory of classical gravity coupled to quantum matter fields with  fundamentally stochastic dynamics \cite{O18}.

Similarly, our theorem can be understood as shedding light about possible limitations that QT might have if spacetime were fundamentally classical. For example, violation of our assumptions \ref{assu5} or \ref{assu7} suggest that it might not be possible to maintain the location of a sufficiently massive system in quantum superposition beyond a certain time, or that it might not be possible to distribute quantum entanglement beyond a certain distance. Alternatively, the spacetime measurement $\mathcal{SM}$ in our thought experiment might require different dynamics to the ones discussed in assumption \ref{assu6}.

However, assumption \ref{assu1} in our theorem is very strong. Thus, a standard viewpoint could simply be that this assumption cannot hold due to apparently being in conflict with general relativity. Nevertheless, we have argued that this assumption does not violate relativistic causality. In particular, the satisfaction of the no-signalling principle by quantum theory implies that a superluminal quantum state reduction, as suggested in assumption \ref{assu1}, does not allow us to communicate information faster than light. Moreover, strictly speaking, general relativity does not say how spacetime should be described in the presence of matter in quantum superpositions or in entangled quantum states. It is plausible that some notions about relativistic causality might need extension in a theory unifying quantum theory and general relativity. \edit{Assumption \ref{assu1} is supported by the observed violation of Bell inequalities \cite{HBDRKBRVSAAPMMTEWTH15,GVMWHHPSKLAAPMBGLSNSUWZ15,SMCBWSGGHACDHLVLTMZSAAPJMKBMKN15}, if we assume that QSR is a real physical process.} 
Nonetheless, it would be interesting to investigate further arguments that do not require this arguable assumption.

Although our other assumptions, assumptions \ref{free}--\ref{minkowski} and assumption \ref{assu3}, might seem more reasonable, it is in principle possible that they could need re-evaluation in a unification of general relativity with quantum theory. For example, as mentioned, Maldacena and Susskind \cite{MS13} have speculated that quantum systems in entangled states might be connected via wormholes.

Our theorem assumes ideal situations with no errors. However, it can be straightforwardly extended
to allow for small error probabilities. In particular, assumption \ref{assu3} considers that there are classical spacetime geometries that can be perfectly
distinguished. This can be extended to allow for a small probability of error. Assumption \ref{assu5} considers that the quantum superposition
for the location of a massive system can be maintained perfectly for a sufficiently long time. This assumption can be relaxed so that
the change of fidelity of the quantum state for this superposition can be kept within a small range for the considered time interval. Finally,
assumption \ref{assu7} considers the distribution of a perfect singlet state over long distances. This can be extended to allow the distribution
of a quantum entangled state that is close to a singlet.

Bell's theorem \cite{Bell}, that there exist quantum correlation violating local causality, is one of the most striking features of quantum theory.
For this reason, we have \edit{argued} here that the satisfaction of Bell inequalities by spacetime degrees of freedom in a background Minkowski spacetime
is a necessary condition for spacetime to be sensibly called ``classical". However, this might change with different definitions of classicality.
Related to this, we note that Kent \cite{K09} has proposed a definition of the non-local causality of spacetime. According to Kent's definition, spacetime is non-locally causal if the measurement choices and outcomes of a Bell experiment on distant entangled quantum particles can be amplified
with different macroscopic gravitational fields whose measurement outcomes violate a Bell inequality.

\begin{acknowledgments}
The author acknowledges financial support from the UK Quantum
Communications Hub grant no. EP/T001011/1 and thanks Adrian Kent for helpful conversations.
\end{acknowledgments}

\appendix
\section{Proof that the probability distribution (\ref{newa5.2}) violates the CHSH inequality}
\label{appchsh}

We show that the CHSH inequality (\ref{chsh}) is violated, and the Tsirelson bound (\ref{cirel}) is saturated, by the probability distribution $P_{AS}(as\vert \alpha1\beta 1)$ given by (\ref{newa5.2}), i.e., when Alice sets $\alpha'=1$ and Bob set $\beta'=1$, where
\begin{equation}
\label{app1}
I_\text{CHSH}=E(0,0)+E(0,1)-E(1,0)+E(1,1),
\end{equation}
and where
\begin{equation}
\label{app2}
E(\alpha,\beta)=\sum_{a=0}^1\bigl[P_{AS}(aa\vert\alpha1\beta 1)-P_{AS}(a\bar{a}\vert \alpha1\beta 1)\bigr],
\end{equation}
for all $\alpha,\beta\in\{0,1\}$. 

In the main text, we defined the qubit orthogonal bases $\mathcal{B}_{\alpha \alpha'}=\{\lvert \phi_{\alpha\alpha'}^a\rangle\}_{a=0}^1$, 
where the quantum states $\lvert \phi_{\alpha\alpha'}^a\rangle$ are given by (\ref{e1}), for all $\alpha,\alpha',a\in\{0,1\}$. It follows straightforwardly from (\ref{e1}) that
\begin{equation}
\label{app3}
\lvert \langle \phi_{\alpha 1}^a\vert \phi_{\beta 0}^{\bar{s}}\rangle \rvert^2=\frac{1}{2}\biggl(1-(-1)^{a\oplus s}\frac{1}{\sqrt{2}}\biggr),
\end{equation}
if $(\alpha,\beta)\in\bigl\{(0,0),(0,1),(1,1)\bigr\}$, and that
\begin{equation}
\label{app4}
\lvert \langle \phi_{\alpha 1}^a\vert \phi_{\beta 0}^{\bar{s}}\rangle \rvert^2=\frac{1}{2}\biggl(1+(-1)^{a\oplus s}\frac{1}{\sqrt{2}}\biggr),
\end{equation}
if $(\alpha,\beta)=(1,0)$, for all $a,s\in\{0,1\}$, where `$\oplus$' denotes sum modulo 2. Thus, from (\ref{newa5.2}), and from (\ref{app1}) -- (\ref{app4}), we obtain  
\begin{equation}
\label{app5}
I_\text{CHSH}=-2\sqrt{2},
\end{equation}
which violates the CHSH inequality (\ref{chsh}) and saturates the Tsirelson bound (\ref{cirel}), as claimed.


\section{Proof of (\ref{b2})}
\label{app2}

We show (\ref{b2}) in the case that $\mathcal{SM}$ does not change the quantum state of $BB'$. In this case, we can determine the probability distribution $P_\text{AB}(ab\vert \alpha\alpha'\beta 0)$ by ignoring the action of $\mathcal{SM}$. 
This situation is described purely by quantum theory. Since Alice's actions commute with Bob's ones, we can analyse the situation by considering that Bob implements his actions before Alice.

Suppose that Bob inputs $\beta=0$. From (\ref{e1.1}), the initial quantum state of the joint system $ABB'$ can be expressed by
\begin{eqnarray}
\label{a5.4}
&&\lvert\Psi^{-}\rangle_{AB}\lvert x_\text{init}\rangle_{B'}\nonumber\\
&&\qquad=\frac{1}{\sqrt{2}}\bigl(\lvert \phi_{00}^{0}\rangle\lvert \phi_{00}^{1}\rangle-\lvert \phi_{00}^{1}\rangle\lvert \phi_{00}^{0}\rangle\bigr)_{AB}\lvert x_\text{init}\rangle_{B'}.
\end{eqnarray}
In this case $U_{\text{mic}}^\beta=\mathbb{1}$ is the identity operation. From (\ref{e3}) and (\ref{a5.4}), after Bob applies $U_\text{mac}$ on $BB'$ at the time $t_3$, the system $ABB'$ transforms into
\begin{eqnarray}
\label{a5.5}
&&U_\text{mac}\lvert\Psi^{-}\rangle_{AB}\lvert x_\text{init}\rangle_{B'}\nonumber\\
&&\quad=\frac{1}{\sqrt{2}}\bigl(\lvert \phi_{00}^{0}\rangle_{A}\lvert x_{1}(t_3)\rangle_{BB'}-\lvert \phi_{00}^{1}\rangle_A\lvert x_{0}(t_3)\rangle_{BB'}\bigr).\nonumber\\
\end{eqnarray}
The system $BB'$ follows the unitary evolution given by (\ref{position}) during the time interval $[t_3,t_5]$. Thus, at the time $t_5$, the quantum state of the joint system $ABB'$ is given by
\begin{eqnarray}
\label{a5.51}
&&U_{\text{pos}}(t_5)U_\text{mac}\lvert\Psi^{-}\rangle_{AB}\lvert x_\text{init}\rangle_{B'}\nonumber\\
&&\quad=\frac{1}{\sqrt{2}}\bigl(\lvert \phi_{00}^{0}\rangle_{A}\lvert x_{1}(t_5)\rangle_{BB'}-\lvert \phi_{00}^{1}\rangle_A\lvert x_{0}(t_5)\rangle_{BB'}\bigr).\nonumber\\
\end{eqnarray}
Then, Bob measures $BB'$ in the basis $\mathcal{B}_{\text{mac}}=\{\lvert x_b(t_5)\rangle\}_{b=0}^1$ at the time $t_5$. With probability $\frac{1}{2}$, Bob obtains the outcome $b$, the quantum state of his joint system $BB'$ reduces to $\lvert x_b(t_5)\rangle$, and the quantum state of Alice's qubit $A$ reduces to $\lvert \phi_{00}^{\overline{b}}\rangle$, for all $b\in\{0,1\}$. 
Alice measures $A$ in the basis $\mathcal{B}_{\alpha\alpha'}=\{\lvert \phi_{\alpha\alpha'}^{a}\rangle\}_{a=0}^1$ and obtains outcome $a$ with probability $\bigl\lvert \langle \phi_{\alpha\alpha'}^{a}\vert \phi_{00}^{\overline{b}}\rangle\bigr\rvert^2$, for all $a\in\{0,1\}$. Thus, we have
\begin{equation}
\label{a5.6}
P_\text{AB}(ab\vert \alpha\alpha'00)=\frac{1}{2}\bigl\lvert \langle \phi_{\alpha\alpha'}^{a}\vert \phi_{00}^{\overline{b}}\rangle\bigr\rvert^2,
\end{equation}
for all $a,b,\alpha,\alpha'\in\{0,1\}$, which is (\ref{b2}) for the case $\beta=0$.

Now suppose that  Bob inputs $\beta=1$. From (\ref{e1.1}), the initial quantum state of the joint system $ABB'$ can be expressed by
\begin{eqnarray}
\label{a5.7}
&&\lvert\Psi^{-}\rangle_{AB}\lvert x_\text{init}\rangle_{B'}\nonumber\\
&&\qquad=\frac{1}{\sqrt{2}}\bigl(\lvert \phi_{10}^{0}\rangle\lvert \phi_{10}^{1}\rangle-\lvert \phi_{10}^{1}\rangle\lvert \phi_{10}^{0}\rangle\bigr)_{AB}\lvert x_\text{init}\rangle_{B'}.
\end{eqnarray}
In this case, we have $U_\text{mic}^\beta=U_\text{mic}$. From (\ref{e2}), after Bob applies $U_\text{mic}$ on his qubit $B$, the system $ABB'$ transforms into
\begin{eqnarray}
\label{a5.8}
&&U_\text{mic}\lvert\Psi^{-}\rangle_{AB}\lvert x_\text{init}\rangle_{B'}\nonumber\\
&&\quad=-\frac{1}{\sqrt{2}}\bigl(\lvert \phi_{10}^{0}\rangle\lvert \phi_{00}^{1}\rangle+\lvert \phi_{10}^{1}\rangle\lvert \phi_{00}^{0}\rangle\bigr)_{AB}\lvert x_\text{init}\rangle_{B'}.
\end{eqnarray}
Then, Bob applies $U_\text{mac}$ on his joint system $BB'$ and, at the time $t_3$, the quantum state of $ABB'$ transforms into
\begin{eqnarray}
\label{a5.9}
&&U_\text{mac}U_\text{mic}\lvert\Psi^{-}\rangle_{AB}\lvert x_\text{init}\rangle_{B'}\nonumber\\
&&\quad=-\frac{1}{\sqrt{2}}\bigl(\lvert \phi_{10}^{0}\rangle_{A}\lvert x_1(t_3)\rangle_{BB'}+\lvert \phi_{10}^{1}\rangle_{A}\lvert x_0(t_3)\rangle_{BB'}\bigr),\nonumber\\
\end{eqnarray}
as follows from (\ref{e3}). 
The system $BB'$ follows the unitary evolution given by (\ref{position}) during the time interval $[t_3,t_5]$. Thus, at the time $t_5$, the quantum state of the joint system $ABB'$ is given by
\begin{eqnarray}
\label{a5.9.1}
&&U_\text{pos}(t_5)U_\text{mac}U_\text{mic}\lvert\Psi^{-}\rangle_{AB}\lvert x_\text{init}\rangle_{B'}\nonumber\\
&&\quad=-\frac{1}{\sqrt{2}}\bigl(\lvert \phi_{10}^{0}\rangle_{A}\lvert x_1(t_5)\rangle_{BB'}+\lvert \phi_{10}^{1}\rangle_{A}\lvert x_0(t_5)\rangle_{BB'}\bigr).\nonumber\\
\end{eqnarray}
Then, Bob measures $BB'$ in the basis $\mathcal{B}_{\text{mac}}=\{\lvert x_b(t_5)\rangle\}_{b=0}^1$ at the time $t_5$. With probability $\frac{1}{2}$, Bob obtains the outcome $b$, the quantum state of his joint system $BB'$ reduces to $\lvert x_b(t_5)\rangle$, and the quantum state of Alice's qubit $A$ reduces to $\lvert \phi_{10}^{\overline{b}}\rangle$, for all $b\in\{0,1\}$. Alice measures $A$ in the basis $\mathcal{B}_{\alpha\alpha'}=\{\lvert \phi_{\alpha\alpha'}^{a}\rangle\}_{a=0}^1$ and obtains outcome $a$ with probability $\bigl\lvert \langle \phi_{\alpha\alpha'}^{a}\vert \phi_{10}^{\overline{b}}\rangle\bigr\rvert^2$, for all $a\in\{0,1\}$. Thus, we have
\begin{equation}
\label{a5.10}
P_\text{AB}(ab\vert \alpha\alpha'10)=\frac{1}{2}\bigl\lvert \langle \phi_{\alpha\alpha'}^{a}\vert \phi_{10}^{\overline{b}}\rangle\bigr\rvert^2,
\end{equation}
for all $a,b,\alpha,\alpha'\in\{0,1\}$, which is (\ref{b2}) for the case $\beta=1$.

%

\end{document}